\documentclass[
	final, %
	british, %
	orivec,envcountsect,runningheads, %
	a4paper %
]{llncs}
\usepackage{xparse}
\usepackage{etoolbox}
\usepackage{environ}
\usepackage[final]{graphicx}
\usepackage{realboxes}

\makeatletter
\newcommand{\IfDefTF}[3]{\@ifundefined{#1}{#3}{#2}}
\newcommand{\IfDefT}[2]{\IfDefTF{#1}{#2}{}}
\newcommand{\IfDefF}[2]{\IfDefTF{#1}{}{#2}}
\newcommand{\IfLabelExistsTF}[3]{\IfDefTF{r@#1}{#3}{#2}}
\newcommand{\IfLabelExistsT}[2]{\IfLabelExistsTF{#1}{#2}{}}
\newcommand{\IfLabelExistsF}[2]{\IfLabelExistsTF{#1}{}{#2}}
\makeatother

\usepackage[T1]{fontenc}
\usepackage{lmodern}
\usepackage[utf8]{inputenc}
\usepackage{microtype}
\usepackage{inconsolata}

\usepackage{xcolor}

\newcommand{\reducedstrut}{\vrule width 0pt height .9\ht\strutbox depth .9\dp\strutbox\relax}
\newcommand{\zcolorbox}[2]{%
  \begingroup
  \setlength{\fboxsep}{0pt}%
  \colorbox{#1}{\reducedstrut#2\/}%
  \endgroup
}

\colorlet{lstbackground}{black!05}
\colorlet{keyword}{blue!40!black}
\colorlet{literal}{red!40!black}
\colorlet{type}{green!40!black}
\colorlet{comment}{black!20}

\usepackage[british]{babel}

\usepackage{csquotes}
\usepackage[all]{foreign}

\defasforeign[etsimilia]{et similia}
\defasforeign[Etsimilia]{Et similia}
\defnotforeign[wrt]{{\xperiodafter{w.r.t}}}
\defnotforeign[wl]{{\xperiodafter{w.l.o.g}}}

\usepackage{amsmath}
\usepackage{amsfonts}
\usepackage{amssymb}
\usepackage{stmaryrd}
\usepackage{centernot}
\usepackage{mathtools}
\usepackage{esvect}

\allowdisplaybreaks

\usepackage{float}
\usepackage[section]{placeins}

\usepackage[inline]{enumitem}

\usepackage{array}
\usepackage{xtab}
\usepackage{booktabs}

\usepackage{url}
\PassOptionsToPackage{hyphens}{url}
\usepackage[breaklinks=true]{hyperref}
\usepackage{nameref}

\usepackage[htt]{hyphenat}
\def\url#1{\href{#1}{\texttt{#1}}}

\usepackage[capitalise,nameinlink,noabbrev]{cleveref}
\makeatletter
\newcommand{\customlabel}[4][0]{%
  \protected@write\@auxout{}{\string\newlabel{#3}{{#4}{\thepage}{#4}{#3}{}}}%
  \protected@write\@auxout{}{\string\newlabel{#3@cref}{{[#2][#1][#1]#4}{\thepage}}}%
}
\newcommand{\crefv}[1]{%
  \begingroup\@cref@compressfalse\@cref@sortfalse\cref{#1}\endgroup%
}
\newcommand{\Crefv}[1]{%
  \begingroup\@cref@compressfalse\@cref@sortfalse\Cref{#1}\endgroup%
}
\newcommand{\crefabbrev}[1]{%
  \begingroup\@cref@abbrevtrue\cref{#1}\endgroup%
}
\NewDocumentCommand\declarecrefname{s m m m m m}{
  \Crefname{#2}{#5}{#6}
  \IfBooleanTF{#1}{
    \crefname{#2}{#3}{#4}
  }{
    \if@cref@capitalise
      \crefname{#2}{#5}{#6}
    \else
      \crefname{#2}{#3}{#4}
    \fi
  }
}

\makeatother

\AtEndPreamble{\hypersetup{
  hidelinks,
  final,
}}

 \usepackage{proof-dashed}

\NewDocumentCommand{\rname}{o m}{%
  \ensuremath{\left\lfloor\mspace{-3mu}%
  \IfNoValueF{#1}{\textnormal{\scshape {#1}}\middle|}%
  \textnormal{\scshape {#2}}%
  \mspace{-1mu}\right\rceil}}
\newcommand{\rlabel}[2]{{%
  \customlabel{rule}{#2}{#1}%
  \hypertarget{#2}{#1}}}

\NewDocumentCommand{\infrule}{o o m m}{%
    \infer[\IfNoValueF{#1}{\hspace{-2pt}\raisebox{.1ex}{\scalebox{.8}%
        {\IfNoValueTF{#2}{#1}{\rlabel{#1}{#2}}}}}]%
    {#3}{#4}%
}
\crefname{rule}{rule}{rules}
\Crefname{rule}{Rule}{Rules}
\spnewtheorem*{convention}{Convention}{\bfseries}{\normalfont}
\spnewtheorem*{notation}{Notation}{\bfseries}{\normalfont}
\spnewtheorem{setting}{Setting}{\bfseries}{\normalfont}

\crefname{proof}{proof of}{proofs of}
\Crefname{proof}{Proof of}{Proofs of}

\usepackage[
  sectionbib,
  square,
  numbers,
  sort&compress
  ]{natbib}
\usepackage[
  textsize=footnotesize,
  colorinlistoftodos,
  prependcaption
]{todonotes}

\usepackage[final]{listings}

\expandafter\patchcmd\csname \string\lstinline\endcsname{%
  \leavevmode
  \bgroup
}{%
  \leavevmode
  \ifmmode\hbox\fi
  \bgroup
}{}{%
  \typeout{Patching of \string\lstinline\space failed!}%
}

\lstdefinelanguage{Lambda}{%
  morekeywords={%
    if,then,else,fix,let,rec,async,in,fst,snd,take,set %
  },%
  morekeywords={[2]int,[2]bool,[2]string},   %
  otherkeywords={:,=}, %
  literate={%
    {->}{{$\to$}}{2}
    {|->}{{$\mapsto$}}{2}
    {\\}{{$\lambda$}}{1}
  },
  keepspaces,
  mathescape
}[keywords,comments,strings]%
\newcommand\code[2][~]{\lstinline#1#2#1}

\def\undeff#1#2{\undefk(#1,#2)}
\def\undefk{\texttt{undef}}

\def\fn{\mathop{\mathrm{fn}}}
\def\nil{\texttt{0}}
\def\pp{\mathbin{\texttt{|}}}
\def\res#1#2{\mathop{\nu#1}#2}

\def\unit{\code{()}}
\def\fundef#1#2{#1 \mathbin{\texttt{\raisebox{.75pt}{:}\kern-1pt=}} #2}
\def\hnddef#1#2{#1 \mathbin{\texttt{\raisebox{.75pt}{-}\kern-1pt>}} #2}
\def\evalp#1#2{#1 \blacktriangleleft #2}

\def\evalctx{\mathcal{E}}
\let\env\mathcal
\let\congr\equiv
\newcommand\lbox[2][1em]{
	\sbox0{$\scriptstyle #2$}
	\ifdim \wd0 < #1%
		\mathmakebox[#1]{\usebox0}%
	\else%
		\mathmakebox[\wd0]{\usebox0}%
	\fi%
}
\newcommand\reducesto[1][]{\xrightarrow{\lbox{#1}}}
\newcommand\plat[2]{\langle #1,#2 \rangle}

\let\defeq\triangleq
\def\fn{\mathop{\mathrm{fn}}}

\newcommand\scalemath[2]{\scalebox{#1}{\mbox{\ensuremath{\displaystyle #2}}}}

\newcommand{\SKC}{SKC}

\newcommand{\hl}[1]{\zcolorbox{lstbackground}{#1}}
 
\begin{document}

\mainmatter

\title{No more, no less}
\subtitle{A formal model for serverless computing}
\titlerunning{A formal model for serverless computing}

\author{
	Maurizio Gabbrielli\inst{1} \and
	Saverio Giallorenzo\inst{2} \and
	Ivan Lanese\inst{1} \and
	Fabrizio Montesi\inst{2} \and
	Marco Peressotti\inst{2} \and
	Stefano Pio Zingaro\inst{1}
}
\authorrunning{Gabbrielli et al.}

\institute{
	INRIA, France / Università di Bologna, Italy\\
	\email{\{maurizio.gabbrielli,ivan.lanese,stefanopio.zingaro\}@unibo.it}
	\and
	University of Southern Denmark, Denmark\\
	\email{\{saverio,fmontesi,peressotti\}@imada.sdu.dk}
}

\maketitle

\begin{abstract}
Serverless computing, also known as Functions-as-a-Service, is a recent
paradigm aimed at simplifying the programming of cloud applications. The idea
is that developers design applications in terms of functions, which are then
deployed on a cloud infrastructure. The infrastructure takes care of
executing the functions whenever requested by remote clients, dealing
automatically with distribution and scaling with respect to inbound traffic.

While vendors already support a variety of programming languages for
serverless computing (\eg Go, Java, Javascript, Python), as far as we
know there is no reference model yet to formally reason on this
paradigm.
In this paper, we propose the first core formal programming model for serverless
computing, which combines ideas from both the $\lambda$-calculus (for
functions) and the $\pi$-calculus (for communication). To illustrate our
proposal, we model a real-world serverless system. Thanks to our model, we
capture limitations of current vendors and formalise possible amendments.
\end{abstract}

\section{Introduction}
\label{sec:introduction}

Serverless computing~\cite{Jonas2019}, also known as Functions-as-a-Service,
narrows the development of Cloud applications to the definition and
composition of stateless functions, while the provider handles the deployment,
scaling, and balancing of the host infrastructure.
Hence, although a bit of a misnomer --- as servers are of course involved ---
the ``less'' in serverless refers to the removal of some server-related
concerns, namely, their \emph{maintenance}, \emph{scaling}, and expenses
related to a sub-optimal management (\eg idle servers).
Essentially, serverless pushes to the extreme the per-usage model of Cloud
Computing: in serverless, users pay only for the computing resources used at
each function invocation. This is why recent
reports~\cite{HellersteinFGSS19,Jonas2019} address serverless computing as the
actual realisation of the long-standing promise of the Cloud to deliver
\emph{computation as a commodity}.
AWS Lambda~\cite{aws_lambda}, launched in 2014, is the first and most
widely-used serverless implementation, however many players like Google,
Microsoft, Apache, IBM, and also open-source communities recently joined the
serverless
market~\cite{apache_openwhisk,microsoft_azure_functions,google_cloud_functions,ironio_ironFunctions,openlambda,ibm_functions}.
Current serverless proposals support the definition of functions --- written
in mainstream languages such as Go, Java, Javascript or Python --- activated
by specific events in the system, like a user request to a web gateway, the
delivery of content from a message broker or a notification from a database.
The serverless infrastructure transparently handles the instantiation of
functions, as well as monitoring, logging, and fault tolerance.

Serverless offerings have become more and more common, yet the technology is
still in its infancy and presents
limitations~\cite{baldini2017serverless,Jonas2019,HellersteinFGSS19} which
hinder its wide adoption. For example, current serverless implementations
favour operational flexibility (asynchrony and scalability) over developer
control (function composition).
Concretely, they do not support the direct composition of functions, which must
call some stateful service in the infrastructure (\eg a message broker) which
will take care of triggering an event bound to the callee. On the one hand,
that limitation is beneficial, since programmers must develop their functions
as highly fine-grained, re-usable components (reminiscent of service-oriented
architectures and microservices~\cite{DragoniGLMMMS17}). On the other hand,
such openness and fine granularity increases the complexity of the system:
programmers cannot assume sequential consistency or serialisability among their
functions, which complicates reasoning on the semantics of the transformations
applied to the global state of their architecture. This holds true also when
estimating resource usage/costs, due to the complexity of unfolding all
possible concurrent computations.

The above criticisms pushed us to investigate a core calculus for
serverless computing, to reason on the paradigm, to model desirable features of future
implementations, and to formalise guarantees over programs. In
\cref{sec:calculus} we introduce the Serverless Kernel Calculus (\SKC); as far
as we know, the first core formal model for serverless computing. \SKC\ combines ideas
from both the $\lambda$-calculus (for functions) and the $\pi$-calculus (for
communication). In \cref{sec:calculus}, we also extend \SKC{} to capture
limitations of current serverless implementations. In \cref{sec:example} we use
our extension to model a real-world serverless architecture~\cite{tailor},
implemented on AWS Lambda. Finally, in \cref{sec:discussion} we discuss future
developments of \SKC{}.

\section{A Serverless Kernel Calculus}
\label{sec:calculus}

\looseness=-1
Our kernel calculus defines a serverless architecture as a pair
$\plat{S} {\env D}$, where $S$ is the system of \emph{running functions} and
$\env D$ is a \emph{definition repository}, containing function
definitions. The repository $\env D$ is a partial function from function
names $\code f$ to function bodies $M$.
$M$ includes function application (\code{$M$\ $M$'}), asynchronous execution of
new functions (\code{async\ $M$}), function names \code{f}, and values $V$.
Values include variables \code{x}, $\lambda$-abstractions
\code{\\x.$M$}, named
\emph{futures}~\cite{baker1977,halstead1985,NiehrenSS06} \code{$c$}, and the
unit value \unit. A system $S$ contains \emph{running functions} $\evalp{c}
{M}$, where $c$ will contain the result of the computation of the function
$M$. Systems can be composed in parallel $\pp$ and include the empty system
$\nil$. Futures can be restricted in systems via $\res{c} {S}$.%
\begin{align*}
S, S' \Coloneqq {} &
  \evalp{c}{M} \mid
  S \pp S' \mid
  \res{c}{S} \mid
  \nil
  &
  \mbox{(Systems)}
\\
M,M' \Coloneqq {} &
  \code{$M$\ $M$'} \mid
  \code{async\ $M$} \mid
  \code{f} \mid
  V 
  &
\mbox{(Functions)}
\\
V,V' \Coloneqq {} &
  \code{x} \mid
  \code{\\x.$M$} \mid
  \code{$c$} \mid
  \unit &
\mbox{(Values)}
\end{align*}
We assume futures to appear only at runtime and not in initial systems.
Moreover, we consider a standard structural congruence $\congr$ that supports
changing the scope of restrictions to avoid name capture, and where parallel
composition is associative, commutative, and has $\nil$ as neutral element.
\begin{align*}
\res{c}{\res{c'}{S}} \congr \res{c'}{\res{c}{S}}
\qquad
\res{c}{(S \pp S')} \congr \res{c}{S} \pp S' \qquad\text{if }  c \not\in \fn(S')
\\
S \congr S \pp \nil
\qquad 
S \pp S' \congr S' \pp S 
\qquad 
(S \pp S') \pp S'' \congr S \pp (S' \pp S'') 
\end{align*}
We define the semantics of our calculus using evaluation contexts
$\evalctx$ and $\evalctx_{\lambda}$, to evaluate, respectively,
systems and functions.
\begin{align*}
	\evalctx {} \Coloneqq
	       \evalp{c}{\evalctx_{\lambda}}
   \hspace{5em}
	\evalctx_{\lambda} {} \Coloneqq
		     [-]
 		\mid \code{(\\x.$M$)$\evalctx_{\lambda}$}
		\mid \code{$\evalctx_{\lambda}$ $M$}
\end{align*}
\looseness=-1
We report in \cref{fig:reduction} the semantics of serverless architectures $
\plat{S} {\env D}$, expressed as reduction rules. Rule \rname{$\beta$} is the
traditional function application of $\lambda$-calculus. Rule
\rname{ret} retrieves the body of function \code{f} from the definition
repository $\env{D}$. Rule \rname{async} models the execution of new
functions: it creates a fresh future $c$ and, in parallel, it
executes function $M$ so that $c$ will store the evaluation of
$M$. When the evaluation of a function reduces to a value, rule
\rname{push} returns the value to the associated future and removes
both the terminated function and its restriction. Rules
\rname{str}, \rname{res}, and \rname{lpar} perform the closure under,
respectively, structural congruence, restriction, and parallel
composition.
\begin{figure}[t]
\begin{spreadlines}{2pt}
\begin{gather*}
\scalemath{.88}{
\infer[\rname{$\beta$}]
{\plat{\evalctx[\code{(\\x.$M$)\ $V$}]}{\env D}
\reducesto
\plat{\evalctx[M\{V/\code{x}\}]}{\env D}}
{}
}
\qquad
\scalemath{.88}{
\infer[\rname{ret}]
{\plat{\evalctx[\code{f}]}{\env D} \reducesto \plat{\evalctx[M]}{\env D}}
{\env D(\code{f})=M}
}
\\
\scalemath{.88}{
\infer[\rname{async}]{
\plat{\evalctx[\code{async\ $M$}]}{\env D}
\reducesto
\plat{\res{c}{(\evalctx[c] \pp \evalp{c}{\code{$M$}})}}{\env D}
}{c \notin \fn(M)}
}
\quad
\scalemath{.88}{
\infer[\rname{push}]
{\plat{\res{c}{(S \pp \evalp{c}{V})}}{\env D} \reducesto \plat{S\{V/c\}}{\env D}}
{}
}
\\
\scalemath{.88}{
  \infer[\rname{str}]
	{\plat{S_0}{\env D} \reducesto \plat{S_1}{\env D'}}
	{S_0 \congr S_0'
	&\plat{S_0'}{\env D} \reducesto \plat{S_1'}{\env D'}
	&S_1' \congr S_1}
}
\\
\scalemath{.88}{
\infer[\rname{res}]
	{\plat{\res{c}{S}}{\env D} \reducesto \plat{\res{c}{S'}}{\env D'}}
	{\plat{S}{\env D} \reducesto \plat{S'}{\env D'}}
}
\qquad
\scalemath{.88}{
\infer[\rname{lpar}]
	{\plat{S_1 \pp S_2}{\env D} \reducesto \plat{S_1' \pp S_2}{\env D'}}
	{\plat{S_1}{\env D} \reducesto \plat{S_1'}{\env D'}}
}
\end{gather*}
\end{spreadlines}
\caption{\SKC{} reduction semantics.}\label{fig:reduction}
\end{figure}
We include in \SKC{} standard components (conditionals, \etc) and extend evaluation contexts ($\evalctx$) accordingly:
\begin{gather*}
M \Coloneqq {} \dots \mid \code{if {}$M$ then {}$M$' else {}$M$''} 
 \mid \code{fst {}$M$} \mid \code{snd {}$M$}
\\
V \Coloneqq {} \dots \mid \code{True} \mid \code{False} \mid \code{
($V$,$V$')}
\end{gather*}
We define standard macros for \code{fix}point, \code{let} and \code{let} 
\code{rec} declarations, and pairs.
\begin{gather*}
\code{fix} \defeq \code{\\f.(\\x.f(xx))(\\x.f(xx))}
\qquad
\code{let x =\ $M$ in\ $M$'} \defeq \code{(\\x.$M$')\ $M$}
\\
\code{let rec x =\ $M$ in\ $M$'} \defeq 
\code{let x =\ fix \\x.$M$ in\ $M$'}
\\
\code{\\(x,y).M} \defeq \code{\\z.(\\x.\\y.M) (fst z) (snd z)}
\end{gather*}

\subsection{\SKC{}$_\sigma$ - A stateful extension of \SKC{}}
\label{sub:stateful_skc}

\SKC{} considers static definition repositories, \ie no rules mutate the state
of $\env{D}$. We now present \SKC{}$_\sigma$, an extension of \SKC{} which
includes two primitives to define transformations on definition repositories.
As shown in \cref{sec:example}, \SKC{}$_\sigma$ is powerful enough to encode
stateful services, like databases and message queues.
\begin{align*}
M,M' \Coloneqq {} & \dots \mid \code{set f\ $M$}  \mid \code{take f}
\end{align*}
The first primitive included in \SKC{}$_\sigma$ is \code{set f\ $M$}, which
updates the definition repository $\env{D}$ to map \code{f} to $M$: users can
use the \code{set} primitive to deploy new function definitions or
update/override existing ones. The second primitive is \code{take f}, which
removes the definition of \code{f} from $\env{D}$, returning it to the
caller. We report below the semantics of the new primitives.
\begin{spreadlines}{1pt}
\begin{gather*}
\scalemath{.88}{\infer[\rname{set}]
{\plat{\evalctx[\code{set f {}$M$}]}{\env{D}}
\reducesto
\plat{\evalctx[\code{f}]}{\env{D}[\code{f|-> }$M$]}}
{\texttt{futures}(M) = \emptyset}}
\\
\scalemath{.88}{\infer[\rname{take}]
{\plat{\evalctx[\code{take f}]}{\env{D}}
\reducesto
\plat{\evalctx[\code{let rec f=$M$ in\ $M$}]}{\undeff{\env{D}}{\code{f}}}}
{\env{D}(\code{f}) = M}}
\end{gather*}
\end{spreadlines}
The only restriction on the application of rule \rname{set} is that the body
$M$ of the newly deployed function $\code f$ does not contain futures
(\texttt{futures}($M$) is the set of futures occurring in $M$). This preserves
the semantics of restriction of futures in function evaluations (\cf rules
\rname{async} and \rname{push}). In the reductum, the rule returns the name of
the deployed function, useful to invoke it in the continuation.
Rule \rname{take} removes the definition $M$ of a deployed function $
\code f$. For simplicity, we define $\rname{take}$ applicable only if
$\code f$ is defined. In the reductum, the caller of the \code{take} obtains
the \code{rec}ursive \code{let} declaration of the function (useful for
internal application) while the association for $\code f$ is removed from $\env D$ by function $\undefk$.

\subsection{\SKC{}$_{\mathtt e}$ - Event-based function composition in \SKC{}}
\label{sub:events_skc}
We present an idiom of \SKC{}, called \SKC{}$_{\mathtt e}$, which
models event-based function composition.  \SKC{}$_{\mathtt e}$
captures one of the main limitations of current serverless vendors:
the lack of support for direct function invocation, replaced by an
event-handling/event-triggering invocation model. Indeed, current
serverless implementations, such as AWS Lambda, work as follows: they
include
 infrastructural stateful services, such as API
gateways, that we can model using our stateful extension
\SKC{}$_{\mathtt \sigma}$, and these services throw
events. User-defined functions are invoked as handlers of these
events. User-defined functions can then invoke the infrastructural
services above. Notably, a user-defined function cannot directly
invoke another user-defined function. We will see an instance of the
event-based pattern in \Cref{sec:example}, while we describe below
event handling mechanisms.

We model events (\code{e} and variations thereof) inside \SKC{} as function names associated
with peculiar function bodies in the repository $\env{D}$ that asynchronously
evaluate the corresponding event handler and discard the handler result. For
convenience, \emph{i}) we package the asynchronous call of an event handler in
the helper function \code{callHandler} below (hereafter, we assume that
$\env{D}$ contains \code{callHandler}) and \emph{ii}) we write \code{_} for
unused variable symbols in binding constructs.
\[
	\code{callHandler} \mapsto \code{\\h.\\x.let _ = async (h () x) in ()}
\]
Event \code{e} is defined in $\env D$ as $\code{e} \mapsto \code{callHandler \\_.h$\textsubscript{e}$}$ and its event handler as $\code{h$_\text{e}$} \mapsto \code{$M_\text{e}$}$; we wrap the name $\code{h$_\text{e}$}$ in a lambda abstraction to avoid expansion (via \rname{Ret}) since function names are not values.
Raising an event \code{e} with some parameter \code{v} results in
asynchronously executing the corresponding handler, as shown by the derivation
below (we abbreviate $\plat{S}{\env{D}}\reducesto \plat{S'}{\env{D}}$ as  
$S \reducesto_{\env{D}} S'$ and label reductions with the
names of the most relevant applied rules).
\begin{spreadlines}{0em}
\[\scalemath{.88}{\begin{array}{l}
\hl{\code{e v}}
\reducesto[\rname{ret}]_{\env{D}}
\hl{\code{callHandler \\_.h$_\text{e}$ v}}
\reducesto[\rname{ret}]_{\env{D}}
\hl{\code{(\\h.\\x.let _=async (h () x) in ()) \\_.h$_\text{e}$ v}}
\\
\reducesto[\rname{$\beta$},\rname{$\beta$}]_{\env{D}} 
\hl{\code{let _=async (\\_.h$_e$ () v) in ()}} %
\reducesto[\rname{async}]_{\env{D}} 
\hl{$\res{c}{(\code{let _=$c$ in ()} \pp \evalp{c}{\code{\\_.h$_e$ () v}})}$}
\\
\reducesto[\rname{$\beta$},\rname{$\beta$}]_{\env{D}}
\hl{$\res{c}{(\code{()} \pp \evalp{c}{\code{h$_e$ v}})}$}
\reducesto[\rname{ret}]_{\env{D}}
\hl{$\code{()} \pp \res{c}{(\evalp{c}{\code{$M_\text{e}$ v}})}$}  
\end{array}}\]
\end{spreadlines}

\section{An Illustrative Example}
\label{sec:example}

\begin{figure}[tb]
 \centering
  \includegraphics[width=.8\textwidth]{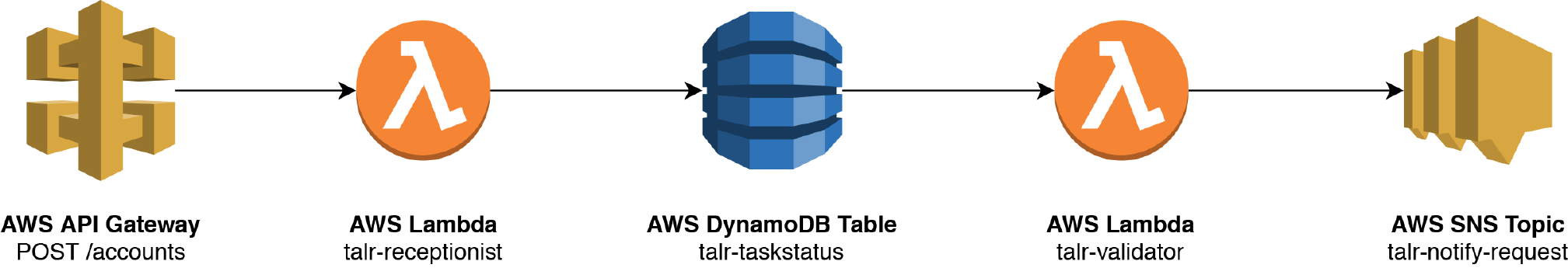}
  \noindent\hfil\rule{0.5\textwidth}{.4pt}\vspace{1em}\hfil
  \includegraphics[width=.8\textwidth]{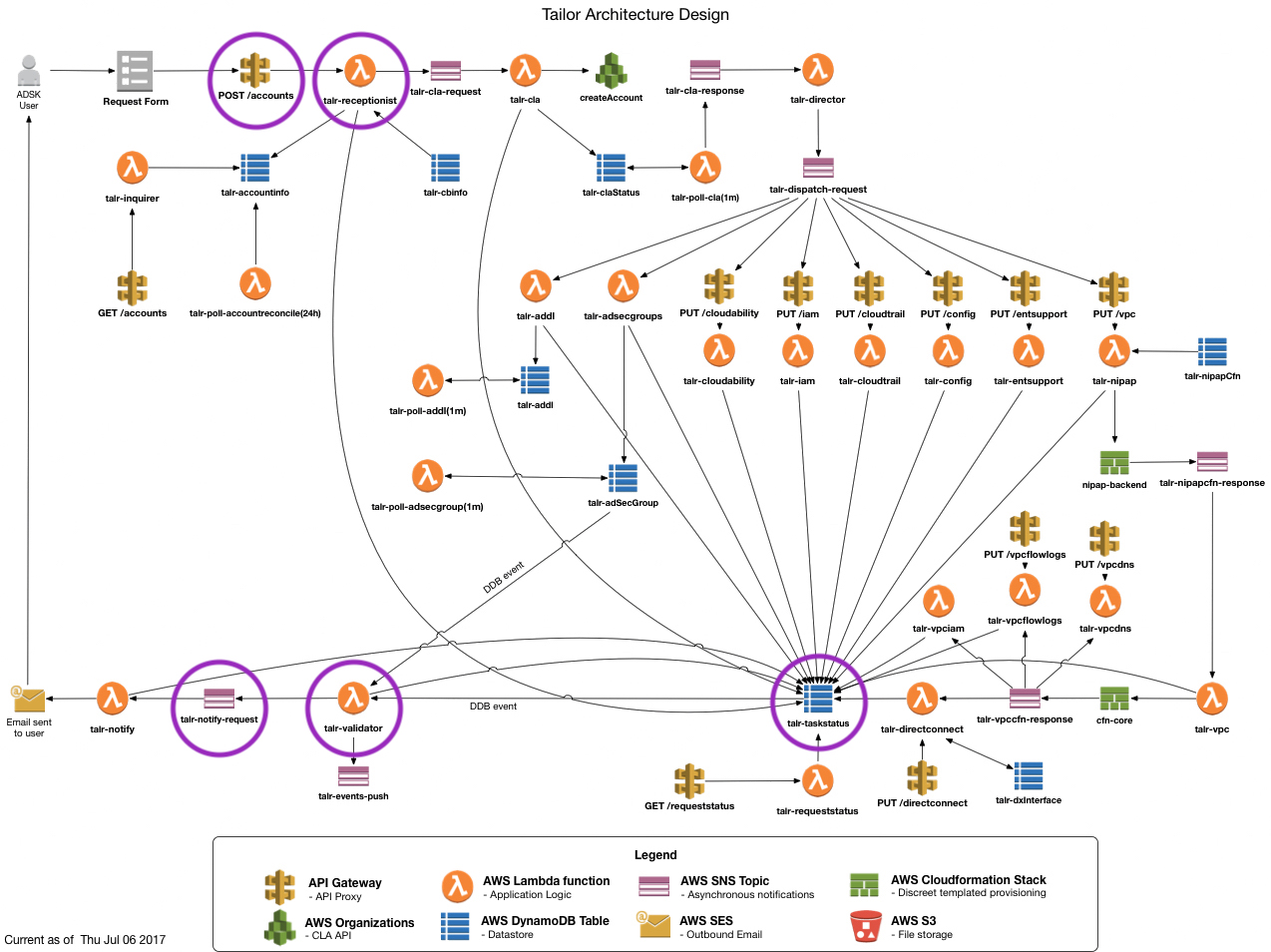}
 \caption{
  Scheme of the Autodesk Tailor system. Top, excerpt
  considered in the example. Bottom, full architecture (circled elements
  belong to the excerpt).}
 \label{fig:tailor}
\end{figure}
Let \SKC{}$_{\sigma \mathtt e}$ be the compound of \SKC{}$_\sigma$ and
\SKC{}$_{\mathtt e}$ presented in \cref{sec:calculus}. Here, we illustrate how
\SKC{}$_{\sigma {\mathtt e}}$ can capture real-world serverless systems by
encoding a relevant portion (depicted in \cref{fig:tailor}) of
Tailor~\cite{tailor}, an architecture for user registration, developed by
Autodesk over AWS Lambda.
Tailor mixes serverless functions with vendor-specific services: \textit{API
Gateways}, key-value databases (\textit{DynamoDB}), and queue-based
notification services (\textit{SNS}). In the architecture, each function
defines a fragment of the logic of a user-registration procedure, like the
initiation of registration requests (\code{talr-receptionist}), request
validation (\code{talr-validator}), \etc.
To model \cref{fig:tailor} in \SKC{}$_{\sigma \code e}$, first, we install in
$\env{D}$ the event handlers for the \textit{API Gateway}, the
\textit{DynamoDB}, and \textit{SNS} services\footnote{We omit the name of the
function called by $\code{e}_\texttt{SNS}$, excluded in the excerpt of
\cref{fig:tailor}.}:
\begin{gather*}
\hl{\code{e$_{\text{API}}$}} \mapsto \code{callHandler(talr-receptionist) }
\qquad
\hl{\code{e$_{\text{SNS}}$}} \mapsto \code{callHandler([...]) }
\\
\hl{\code{e$_{\text{DDB}}$}} \mapsto \code{callHandler(talr-validator)} 
\end{gather*}
Then, we define the functions called by the handlers installed above, using
the same names of the \emph{AWS Lambda} functions in \cref{fig:tailor}.
Handler \code{e$_{\text{API}}$} calls function \code{talr-receptionist}, which
validates the request and inserts the information of the user in the key/value
database. For brevity, we omit the behaviour of \code{talr-receptionist} in
case of invalid requests and the definition of auxiliary functions
\code{validate_request}, \code{get_key}, \code{get_value} in $\env D$:
\begin{align*}
\hl{\code{talr-receptionist}}
\mapsto {} & 
\code{\\x.if validate_request x then}
\\ & 
\quad\code{write_db (get_key x,get_value x) else [...]}
\end{align*}
Handler \code{e$_{\text{DDB}}$} invokes function \code{talr-validator}, which
retrieves from the database the \code{status} of task \code{x}, checks if it is
complete, and sends a notification on \emph{SNS}. We omit the definitions of
functions \code{check} and \code{compose_msg} and of the \code{else} branch.

\begin{align*}
\hl{\code{talr-validator}} \mapsto {} & 
\code{\\x.let status = read_db x in} \\ & 
\code{if check status then push (compose_msg x) else [...]}
\end{align*}
We conclude illustrating the definitions of functions \code{write_db},
\code{read_db}, and \code{push} in $\env D$, which exemplify how
\SKC{}$_{\sigma \mathtt{e}}$ can encode stateful, event-triggering services.
Keys are represented as function names and values are stored in $\env{D}$; thus keys are passed around wrapped in lambda abstractions (\code{\\_.k}) as done for events.
\begin{gather*}
\hl{\code{write_db}} \mapsto \code{\\(x,v).e$_{\text{DDB}}$ (set (x ())
v)}
\quad
\hl{\code{read_db}} \mapsto \code{\\x.x ()}
\\
\hl{\code{push}} \mapsto \code{\\(x,v).e$_{
\text{SNS}}$ (set (x ()) v)}
\end{gather*}
Function \code{write_db} takes a key (wrapped as $\code{x} = \code{\\_.k}$) and a value $\code v$ as parameters, writes on the database by \code{set}ting to $\code v$ the body of a function called $\code k$, and notifies the write, invoking \code{e$_{\text{DDB}}$}\footnote{%
More involved variants of the database are possible. \Eg to avoid clashes
among services using the same key for different elements, we can either use
scoping or prefix key names with service names
--- \eg Tailor uses service-specific tables in \emph{DynamoDB}.
}.
Function \code{read_db} simply unwraps the key thus enabling retrieval from $\env{D}$.
Similarly to \code{write_db}, function \code{push} publishes (\code{set}) a
message $\code v$ on an \textit{SNS} topic (represented as a function name) and triggers \code{e$_{\text{SNS}}$}.

\begin{remark}
The example illustrates how \SKC{} can capture (but not be restricted by) one
of the most prominent limitations of current serverless
platform~\cite{HellersteinFGSS19}, \ie that \emph{i}) user-defined functions
can be only invoked by raising an event that executes a new function (as done
by \code{callHandler}, using the \code{async} primitive) and \emph{ii})
functions can invoke other functions only by interacting with some
event-triggering infrastructural service (\eg a database, represented by 
function \code{write_db}, or a notification queue, represented by function 
\code{push}).
\end{remark}

\section{Discussion and Conclusion}
\label{sec:discussion}
We propose \SKC, the first core formal model to reason on serverless computing.
While the design of \SKC\ strives for minimality, it captures the main
ingredients~\cite{HellersteinFGSS19,Jonas2019} of serverless architectures:
\emph{i}) the deployment and instantiation of event-triggered, stateless
functions and \emph{ii}) the desiderata of direct function-to-function
invocation based on futures --- in \cref{sec:example} we
show how this mechanism is powerful enough to cover also the current setting
of serverless vendors, where function invocation must rely on third-party
services that handle event triggering.

Futures~\cite{baker1977,halstead1985}, which are the main communication
mechanism in \SKC{}, are becoming one of the de-facto standards in
asynchronous systems~\cite{Goetz06,ECMAScript2018,Summerfield13,Williams17}.
We considered using named channels
(as in CCS/$\pi$-calculus~\cite{Milner80,Sangiorgi01}) instead of futures,
but we found them too general for the needs of the serverless model (they are
bi-directional and re-usable). Besides, futures can encode
channels~\cite{NiehrenSS06}.

\looseness=-1
The work closest to ours is~\cite{JPBDMSBG19}, appeared during the submission
of this work, in the form of a technical report. It presents a detailed operational semantics that captures the low-level
details of current serverless implementations (\eg cold/warm components, storage, and transactions are primitive features of their model) whereas \SKC{} identifies a kernel model of serverless computing.
Another work close to SKC is~\cite{NiehrenSS06}, where the authors introduce a
$\lambda$-calculus with futures. Since the aim of~\cite{NiehrenSS06} is to
formalise and reason on a concurrent extension of Standard ML, their calculus
is more involved than \SKC{}, as it contains primitive operators (handlers and
cells) to encode safe non-deterministic concurrent operations, which can be
encoded as macros in \SKC{}. An interesting research direction is to
investigate which results from~\cite{JPBDMSBG19,NiehrenSS06} can be adapted to
\SKC{}.

\looseness=-1
Being the first core framework to reason on serverless architectures, \SKC
{} opens multiple avenues of future research.
For example, current serverless technologies offer little guarantee on
sequential execution across functions, which compels the investigation of new
tools to enforce sequential consistency~\cite{Lamport79} or
serialisability~\cite{Papadimitriou79} of the transformations of the global
state~\cite{HellersteinFGSS19}. That challenge can be tackled developing
static analysis techniques and type disciplines~\cite{HuttelLVCCDMPRT16,AnconaBB0CDGGGH16} for \SKC{}.
Another direction concerns programming models, which should give to programmers
an overview over the overall logic of the distributed functions and capture the
loosely-consistent execution model of serverless~\cite{HellersteinFGSS19}.
Choreographic Programming~\cite{Montesi15,Cruz-FilipeM16} is a promising
candidate for that task, as choreographies are designed to capture the global
interactions in distributed systems~\cite{Kavantzas05}, and recent
results~\cite{CarboneM13,PredaGGLM16,GiallorenzoMG18} confirmed their
applicability to microservices~\cite{DragoniGLMMMS17}, a neighbouring domain
to
that of serverless architectures.  Other possible research directions, that we
do not discuss for space constraints, include monitoring, various kinds of
security analysis including ``self-DDoS
attacks''~\cite{serverless_out_of_control,beware_runonstartup,serverless_a_lesson_learned}
and performance analysis.  This last one is particularly relevant in the
per-usage model of serverless architectures, yet requires to extend \SKC{} with an explicit notion of time in order to support quantitative behavioural reasoning for timed systems \cite{brengos:lmcs2019,brengos:concur2016}.

\phantomsection
\label{structure:end-of-mainmatter}

\bibliography{biblio}

\begin{thebibliography}{37}
\providecommand{\natexlab}[1]{#1}
\providecommand{\url}[1]{\texttt{#1}}
\expandafter\ifx\csname urlstyle\endcsname\relax
  \providecommand{\doi}[1]{doi: #1}\else
  \providecommand{\doi}{doi: \begingroup \urlstyle{rm}\Url}\fi

\bibitem[{Alan Williams}()]{tailor}
{Alan Williams}.
\newblock {Tailor - the AWS Account Provisioning Service}.
\newblock \url{https://github.com/alanwill/aws-tailor}.
\newblock Online; acc. 02/2019.

\bibitem[Ancona et~al.(2016)]{AnconaBB0CDGGGH16}
D.~Ancona et~al.
\newblock Behavioral types in programming languages.
\newblock \emph{Foundations and Trends in Programming Languages}, 3\penalty0
  (2-3):\penalty0 95--230, 2016.

\bibitem[{{Apache}}()]{apache_openwhisk}
{{Apache}}.
\newblock {OpenWhisk}.
\newblock \url{https://github.com/apache/incubator-openwhisk}.
\newblock Online; acc. 02/2019.

\bibitem[{{AWS}}()]{aws_lambda}
{{AWS}}.
\newblock {Lambda}.
\newblock \url{https://aws.amazon.com/lambda/}.
\newblock Online; acc. 02/2019.

\bibitem[Baker~Jr and Hewitt(1977)]{baker1977}
H.~C. Baker~Jr and C.~Hewitt.
\newblock The incremental garbage collection of processes.
\newblock In \emph{ACM Sigplan Notices}, volume 12(8), pages 55--59. ACM, 1977.

\bibitem[Baldini et~al.(2017)]{baldini2017serverless}
I.~Baldini et~al.
\newblock Serverless computing: Current trends and open problems.
\newblock In \emph{Research Advances in Cloud Computing}, pages 1--20.
  Springer, 2017.

\bibitem[Brengos and Peressotti(2016)]{brengos:concur2016}
T.~Brengos and M.~Peressotti.
\newblock A uniform framework for timed automata.
\newblock In \emph{{CONCUR}}, volume~59 of \emph{LIPIcs}, pages 26:1--26:15.
  Schloss Dagstuhl - Leibniz-Zentrum fuer Informatik, 2016.

\bibitem[Brengos and Peressotti(2019)]{brengos:lmcs2019}
T.~Brengos and M.~Peressotti.
\newblock Behavioural equivalences for timed systems.
\newblock \emph{Logical Methods in Computer Science}, 15\penalty0 (1), 2019.

\bibitem[Carbone and Montesi(2013)]{CarboneM13}
M.~Carbone and F.~Montesi.
\newblock Deadlock-freedom-by-design: multiparty asynchronous global
  programming.
\newblock In \emph{{POPL}}, pages 263--274. {ACM}, 2013.

\bibitem[Cruz{-}Filipe and Montesi(2016)]{Cruz-FilipeM16}
L.~Cruz{-}Filipe and F.~Montesi.
\newblock A core model for choreographic programming.
\newblock In \emph{{FACS}}, Lecture Notes in Computer Science, pages 17--35,
  2016.

\bibitem[{Dalla Preda} et~al.(2017)]{PredaGGLM16}
M.~{Dalla Preda} et~al.
\newblock Dynamic choreographies: Theory and implementation.
\newblock \emph{Logical Methods in Computer Science}, 13\penalty0 (2), 2017.

\bibitem[Dragoni et~al.(2017)]{DragoniGLMMMS17}
N.~Dragoni et~al.
\newblock Microservices: Yesterday, today, and tomorrow.
\newblock In \emph{Present and Ulterior Software Engineering.}, pages 195--216.
  Springer, 2017.

\bibitem[ECMAScript()]{ECMAScript2018}
ECMAScript.
\newblock Ecmascript 2018 language specification.
\newblock \url{http://ecma-international.org/ecma-262/9.0/index.html}.
\newblock Online; acc. 02/2019.

\bibitem[Giallorenzo et~al.(2018)Giallorenzo, Montesi, and
  Gabbrielli]{GiallorenzoMG18}
S.~Giallorenzo, F.~Montesi, and M.~Gabbrielli.
\newblock Applied choreographies.
\newblock In \emph{{FORTE}}, pages 21--40. Springer, 2018.

\bibitem[Goetz et~al.(2006)]{Goetz06}
B.~Goetz et~al.
\newblock \emph{Java concurrency in practice}.
\newblock Pearson Education, 2006.

\bibitem[{{Google}}()]{google_cloud_functions}
{{Google}}.
\newblock {Cloud Functions}.
\newblock \url{https://cloud.google.com/functions}.
\newblock Online; acc. 02/2019.

\bibitem[Halstead~Jr(1985)]{halstead1985}
R.~H. Halstead~Jr.
\newblock Multilisp: A language for concurrent symbolic computation.
\newblock \emph{ACM Transactions on Programming Languages and Systems
  (TOPLAS)}, 7\penalty0 (4):\penalty0 501--538, 1985.

\bibitem[Hellerstein et~al.(2019)]{HellersteinFGSS19}
J.~M. Hellerstein et~al.
\newblock Serverless computing: One step forward, two steps back.
\newblock In \emph{{CIDR}}. www.cidrdb.org, 2019.

\bibitem[Hendrickson et~al.(2016)]{openlambda}
S.~Hendrickson et~al.
\newblock {Serverless Computation with OpenLambda}.
\newblock In \emph{{USENIX}}. {USENIX} Association, 2016.

\bibitem[H{\"{u}}ttel et~al.(2016)]{HuttelLVCCDMPRT16}
H.~H{\"{u}}ttel et~al.
\newblock Foundations of session types and behavioural contracts.
\newblock \emph{{ACM} Comput. Surv.}, 49\penalty0 (1):\penalty0 3:1--3:36,
  2016.

\bibitem[{{IBM}}()]{ibm_functions}
{{IBM}}.
\newblock {Cloud Functions}.
\newblock \url{https://www.ibm.com/cloud/functions}.
\newblock Online; acc. 02/2019.

\bibitem[{{Iron.io}}()]{ironio_ironFunctions}
{{Iron.io}}.
\newblock {IronFunctions}.
\newblock \url{https://open.iron.io}.
\newblock Online; acc. 02/2019.

\bibitem[Jangda et~al.(2019)]{JPBDMSBG19}
A.~Jangda et~al.
\newblock Formal foundations of serverless computing.
\newblock \emph{CoRR}, abs/1902.05870, 2019.
\newblock URL \url{http://arxiv.org/abs/1902.05870}.

\bibitem[Jonas et~al.(2019)]{Jonas2019}
E.~Jonas et~al.
\newblock Cloud programming simplified: A berkeley view on serverless
  computing.
\newblock Technical report, EECS Department, University of California,
  Berkeley, Feb 2019.

\bibitem[{{K-Optional Software}}()]{serverless_out_of_control}
{{K-Optional Software}}.
\newblock {Serverless out of Control}.
\newblock \url{https://koptional.com/2019/01/22/serverless-out-of-control/}.
\newblock Online; acc. 02/2019.

\bibitem[Kavantzas et~al.(2005)Kavantzas, Burdett, Ritzinger, and
  Lafon]{Kavantzas05}
N.~Kavantzas, D.~Burdett, G.~Ritzinger, and Y.~Lafon.
\newblock Web services choreography description language version 1.0, {W3C}
  candidate recommendation.
\newblock Technical report, W3C, 2005.
\newblock \url{http://www.w3.org/TR/ws-cdl-10}.

\bibitem[{Kevin Vandenborne}()]{serverless_a_lesson_learned}
{Kevin Vandenborne}.
\newblock {Serverless: A lesson learned. The hard way.}
\newblock
  \url{https://sourcebox.be/blog/2017/08/07/serverless-a-lesson-learned-the-hard-way/}.
\newblock Online; acc. 02/2019.

\bibitem[Lamport(1979)]{Lamport79}
L.~Lamport.
\newblock How to make a multiprocessor computer that correctly executes
  multiprocess programs.
\newblock \emph{IEEE Trans. Comput.}, 28\penalty0 (9):\penalty0 690--691, Sept.
  1979.

\bibitem[{{Microsoft}}()]{microsoft_azure_functions}
{{Microsoft}}.
\newblock {Azure Functions}.
\newblock
  \href{https://azure.microsoft.com/services/functions}{\texttt{https://azure.microsoft.com/services/func\-tions}}.
\newblock Online; acc. 02/2019.

\bibitem[Milner(1980)]{Milner80}
R.~Milner.
\newblock \emph{A Calculus of Communicating Systems}, volume~92 of
  \emph{Lecture Notes in Computer Science}.
\newblock Springer, 1980.

\bibitem[Montesi(2015)]{Montesi15}
F.~Montesi.
\newblock Kickstarting choreographic programming.
\newblock In \emph{{WS-FM/BEAT}}, pages 3--10. Springer, 2015.

\bibitem[Niehren et~al.(2006)Niehren, Schwinghammer, and Smolka]{NiehrenSS06}
J.~Niehren, J.~Schwinghammer, and G.~Smolka.
\newblock A concurrent lambda calculus with futures.
\newblock \emph{Theor. Comput. Sci.}, 364\penalty0 (3):\penalty0 338--356,
  2006.

\bibitem[Papadimitriou(1979)]{Papadimitriou79}
C.~H. Papadimitriou.
\newblock The serializability of concurrent database updates.
\newblock \emph{J. {ACM}}, 26\penalty0 (4):\penalty0 631--653, 1979.

\bibitem[Sangiorgi and Walker(2001)]{Sangiorgi01}
D.~Sangiorgi and D.~Walker.
\newblock \emph{The Pi-Calculus - a theory of mobile processes}.
\newblock Cambridge University Press, 2001.
\newblock ISBN 978-0-521-78177-0.

\bibitem[Summerfield(2013)]{Summerfield13}
M.~Summerfield.
\newblock \emph{Python in practice: create better programs using concurrency,
  libraries, and patterns}.
\newblock Addison-Wesley, 2013.

\bibitem[{Tom Wright}()]{beware_runonstartup}
{Tom Wright}.
\newblock {Beware ``RunOnStartup'' in Azure Functions --- a serverless horror
  story}.
\newblock
  \url{http://blog.tdwright.co.uk/2018/09/06/beware-runonstartup-in-azure-functions-a-serverless-horror-story/}.
\newblock Online; acc. 02/2019.

\bibitem[Williams(2017)]{Williams17}
A.~Williams.
\newblock \emph{C++ concurrency in action}.
\newblock Manning, 2017.

\end{thebibliography}

\end{document}